\documentclass[aps,prl,twocolumn,groupedaddress,showpacs,superscriptaddress]{revtex4-1}

\usepackage{graphicx}
\usepackage{dcolumn}
\usepackage{bm}
\usepackage{units}
\usepackage{float,amsmath,verbatim,amssymb,stmaryrd}
\usepackage{epsfig}
\usepackage{color}

\usepackage[colorinlistoftodos,textsize=tiny]{todonotes}

\begin{document}

\preprint{APS/123-QED}

\title{Efficient Nonthermal Particle Acceleration by the Kink Instability in Relativistic Jets}

\author{E. P. Alves}
\email{epalves@slac.stanford.edu}
\affiliation{
High Energy Density Science Division, SLAC National Accelerator Laboratory, Menlo Park, CA 94025, USA
}

\author{J. Zrake}
\affiliation{
Physics Department and Columbia Astrophysics Laboratory, Columbia University, 538 West 120th Street, New York, NY 10027
}

\author{F. Fiuza}
\email{fiuza@slac.stanford.edu}
\affiliation{
High Energy Density Science Division, SLAC National Accelerator Laboratory, Menlo Park, CA 94025, USA
}

\begin{abstract}
Relativistic magnetized jets from active galaxies are among the most powerful cosmic accelerators, but their particle acceleration mechanisms remain a mystery. We present a new acceleration mechanism associated with the development of the helical kink instability in relativistic jets, which leads to the efficient conversion of the jet's magnetic energy into nonthermal particles. Large-scale three-dimensional \emph{ab initio} simulations reveal that the formation of highly tangled magnetic fields and a large-scale inductive electric field throughout the kink-unstable region promotes rapid energization of the particles. The energy distribution of the accelerated particles develops a well-defined power-law tail extending to the radiation-reaction limited energy in the case of leptons, and to the confinement energy of the jet in the case of ions. When applied to the conditions of well-studied bright knots in jets from active galaxies, this mechanism can account for the spectrum of synchrotron and inverse Compton radiating particles, and offers a viable means of accelerating ultra-high-energy cosmic rays to $10^{20}$ eV.
\end{abstract}

\maketitle

Extragalactic radio jets are powerful outflows of relativistic magnetized plasma emanating from the central regions of active galaxies. These systems (known as active galactic nuclei, or AGNs) are among the most powerful accelerators of charged particles in the cosmos. They contain relativistic electrons and positrons which radiate, via synchrotron and inverse Compton processes, from radio waves to TeV $\gamma$-rays, attaining energies vastly in excess of the thermal mean \cite{Celotti2008}. AGN jets are also candidate sources of ultra-high-energy cosmic rays (UHECRs), whose energies are observed by ground-based detectors to exceed $\unit[10^{20}]{eV}$ \cite{Nagano2000, Abraham2007}. This hypothesis has gained further support with the recent coincident detection of $\gamma$-rays and a high-energy neutrino from blazar TXS 0506+056 \cite{Icecube2018}, which confirms that AGN jets accelerate high-energy cosmic rays.

The specific mechanisms by which relativistic jets accelerate charged particles to such high energies remains a long-standing mystery. Observations of bright knots in AGN jets (\emph{e.g.} the well-studied HST-1 in M87) suggest that efficient particle acceleration may be taking place at distances of \unit[10]{pc} to \unit[1]{kpc} from the black hole central engine. At these distances the jet's energy exists primarily in the form of magnetic fields, observed to possess a tightly wound helical structure \cite{Harris2003}. The bright knots are nearly stationary features, and are commonly interpreted as recollimation shocks associated with the interaction of the jet with the ambient medium \cite{Stawarz2006,Bromberg2009}. Historically, particle energization in these regions has been attributed to diffusive shock acceleration \cite{Bell1978,Blandford87}. However, recent work \cite{Sironi2013a,Bell2018} indicates that shock acceleration is not efficient in relativistic magnetically dominated plasma.

Another possibility is that particles accelerate by feeding on the copious free energy of the jet's internal magnetic field. This energy may be extracted via the development of hydromagnetic instabilities that act on the jet's helical magnetic field structure, the most relevant of which is thought to be the helical kink instability (KI) \cite{Begelman1998,Giannios2006}. Recent global magnetohydrodynamic (MHD) simulations of the launching and propagation of relativistic jets \cite{Tchekhovskoy2016,Bromberg2016,Duran2016} confirm that the KI can indeed be triggered where the jet recollimates, playing an important role in the dissipation of the jet's magnetic energy. They also reveal that the KI operates internally, distorting only the spine of the jet, without disrupting its global morphology or its ability to propagate to larger distances \cite{Porth2015,Bromberg2016}. This is consistent with observations, which indicate that AGN jets remain globally stable over these bright recollimation regions \cite{Harris2003,Stawarz2006}. It remains unknown, however, \emph{if} and \emph{how} the magnetic energy dissipated by the internal KI is channeled into energetic nonthermal particles. This inherently kinetic physics is not captured within the framework of MHD.

\begin{figure*}[t!]
\begin{center}
\includegraphics[width=\textwidth]{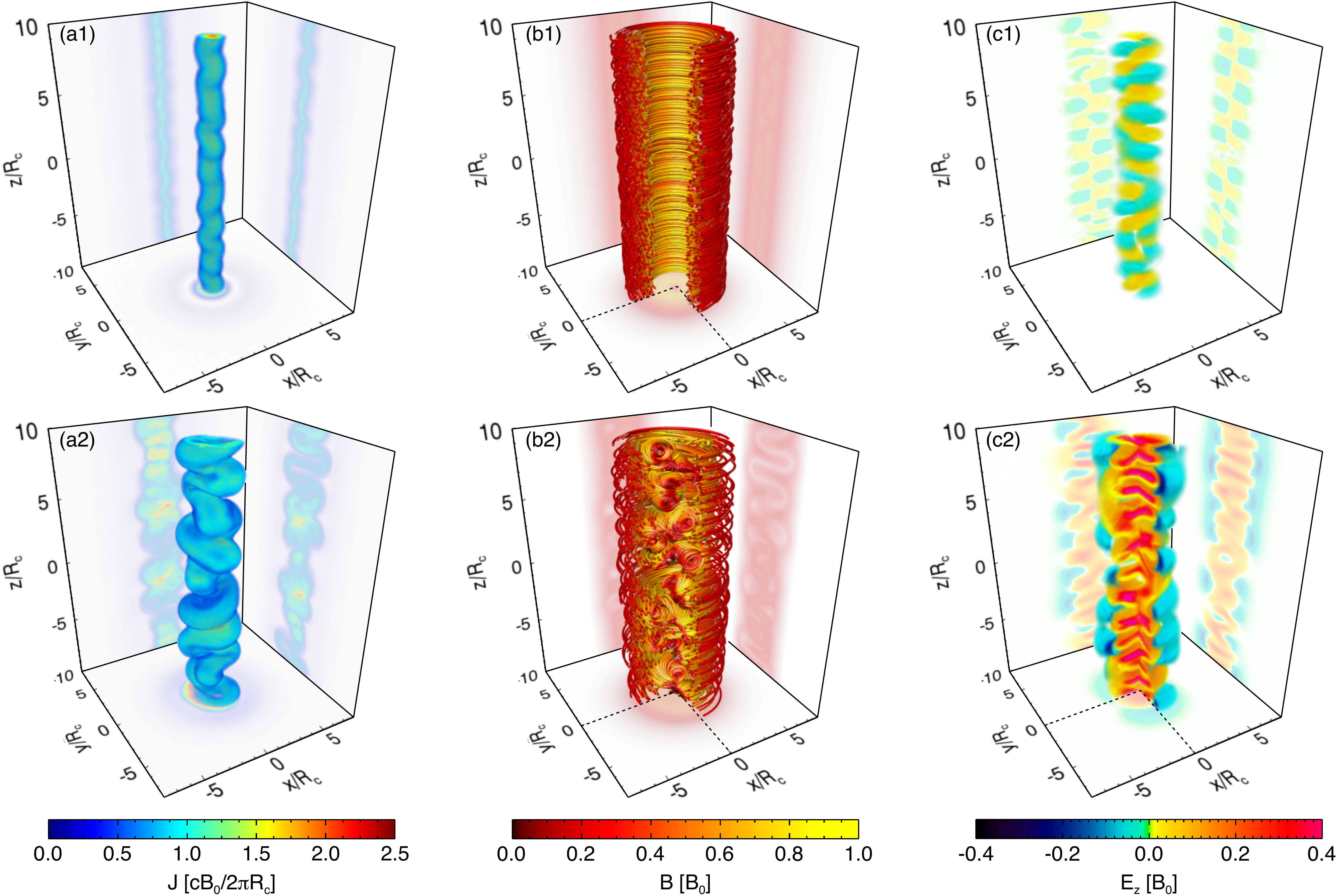}
\caption{Evolution of the jet structure subject to the kink instability.
(a) Current density, (b) magnetic field lines, and (c) axial electric field, taken at times 1) $ct/R_c = 16$ and 2) $ct/R_c = 24$. These times correspond to the linear and nonlinear stages of the kink instability. Note that a quarter of the simulation box has been removed in (b1), (b2), and (c2) to reveal the inner field structure of the jet.
}
\end{center}
\end{figure*}

In this Letter, we present the results of three-dimensional (3D), particle-in-cell (PIC) simulations that reveal for the first time how the self-consistent development of the internal KI in relativistic magnetized jets results in the efficient acceleration of nonthermal particles. We find that the emergence of highly tangled magnetic fields and a large-scale electric field throughout the kink-unstable region promotes rapid energization of the particles up to the confinement energy of the jet. Acceleration occurs over 10 light-crossing times of the jet cross-sectional radius, during which $\sim$50\% of the jet's toroidal magnetic field energy is transfered to newly accelerated particles with a power-law spectrum. Importantly, we observe that the maximum energy gain increases linearly with the jet cross-sectional radius. Based on these findings, we argue that this new mechanism can account for the acceleration of high-energy leptons and hadrons in AGN jets.

We simulated a volume of the jet in its proper reference frame, with relativistic electron-positron plasma supporting a helical magnetic field in an unstable hydromagnetic equilibrium; the net-inward magnetic stress is balanced by increased thermal pressure near the axis (see Supplemental Material \cite{suppl}). This setup approximates the jet spine after the plasma has been focused towards the axis by recollimation, at the moment when it stagnates and is most vulnerable to the internal KI. We consider magnetic field profiles of the form $\mathbf{B}(r) = B_0\frac{r}{R_c}e^{1-r/R_c}\mathbf{e}_\phi + B_z \mathbf{e}_z$, where $R_c$ is the cross-sectional radius of the jet spine. We have also tested toroidal magnetic field profiles that decay as $r^{-\alpha}$ (with $\alpha\geq1$), and determined that our overall findings are not sensitive to the structure of the magnetic field far from $R_c$. Near the black hole, the poloidal and toroidal magnetic field components ($B_z$ and $B_\phi$, respectively) are comparable to one another \cite{Blandford1977}. However, $B_z / B_\phi$ decreases with distance from the source, and can be very small at the relevant $\sim \unit[100]{pc}$ distances. The characteristic magnetic field amplitude (henceforth denoted as $B_0$) at these distances, $B_0 \sim \unit[]{mG}$, is quite strong in the sense that the ratio $\sigma$ of the magnetic to plasma rest-mass energy densities may exceed unity. The simulations cover values of $\sigma = 1 - 10$ and $B_z / B_\phi = 0.0 - 0.5$.

We utilize the fully kinetic electromagnetic PIC code OSIRIS 3.0 \cite{Fonseca2002,Fonseca2008}. Our simulations resolve a large dynamic range in 3D, enabling us to study the interplay between the evolution of the KI at large scales and the dynamics of particles at small scales, i.e. between the MHD physics of the jet spine at $\sim R_c$ and the kinetic physics operating at the particle gyroradius scale $\rho_g \ll R_c$. By systematically increasing the scale separation $\bar{R} \equiv R_c / \langle \rho_g \rangle$, we find asymptotic behavior in the particle acceleration physics as $\bar{R} \gg 1$. The dimensions of the simulated domains are $20\times20\times(10-20)~R_c^3$, with the jet located at the center of the domain and oriented along $\hat{\mathbf{z}}$. The simulations resolve the gyroradius of thermal particles at the core of the jet, $\langle\rho_g\rangle$, with $4-12$ points, and use $8-16$ particles per cell per species \cite{suppl}. Our largest simulations attain $\bar{R} = 50$ and are state-of-the-art in computational scale, following 550 billion particles in $4096 \times 4096 \times 2048$ computational cells.

All of our simulations exhibit qualitatively similar dynamics, illustrated in Figs. 1 and 2 for a jet with $\bar{R} \simeq 8$, $\sigma = 5$, and purely toroidal magnetic field ($B_z = 0$). The KI is triggered by thermal fluctuations in the plasma, and induces a growing helical modulation of the jet spine (top row of Fig. 1) with wavelength $\sim R_c$, consistent with linear theory and MHD simulations \cite{Mizuno2011a}. These transverse motions give rise to an inductive electric field, $\mathbf{E} = -\mathbf{v} \times \mathbf{B}$. At early times, the axial component of this electric field is harmonic, with zero net value along $\mathbf{\hat z}$, $\langle E_z \rangle \simeq 0$ [Fig. 1(c1)]. Notably, we observe that as the instability becomes nonlinear, and the transverse displacements of the jet become comparable to its radius, regions of like-oriented electric field are brought into alignment (Supplemental Fig. S1). This leads to the formation of a coherent inductive electric field throughout the spine region of the jet, with net value $\langle E_z \rangle \simeq 0.2 B_0$ [Fig. 1(c2); a similar $E_z$ field is observed in MHD simulations (not shown here) of the same configuration]. At the same time, strong distortions of the electrical current give rise to a highly tangled magnetic field structure, particularly in the central region where the coherent axial electric field has developed [Fig. 1(b2)]. This configuration of electric and magnetic fields facilitates rapid and efficient transfer of energy from the magnetic field to the plasma particles [Fig. 2(a)], with $\gtrsim 60\%$ of the magnetic energy being dissipated, and the overall plasma magnetization being reduced to $\sim20\%$ of its initial value. This process is completed on a time scale $\tau_{\rm KI} \simeq 10 R_c/c$, which is given by the transit time $2 R_c / v_{\rm KI}$ of the KI-induced transverse motions across the jet diameter, with characteristic speed $v_{\rm KI} \simeq \langle E_z \rangle / B_0 c \simeq 0.2 c$. 

\begin{figure}[t!]
\begin{center}
\includegraphics[width=\linewidth]{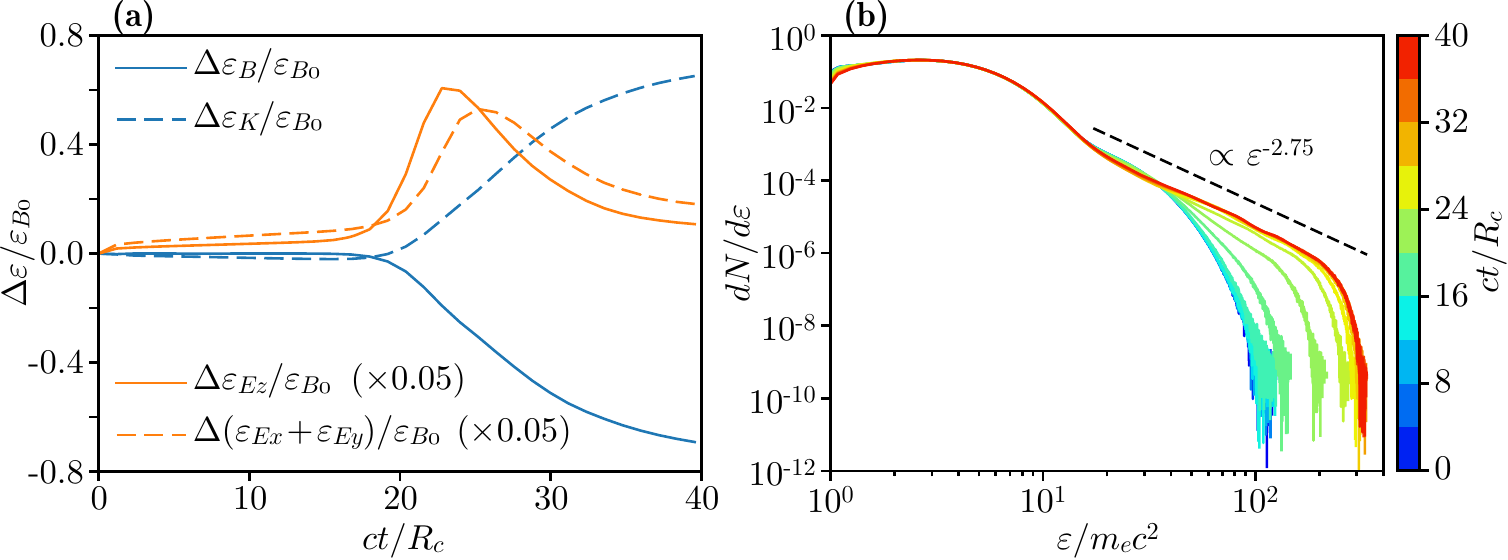}
\caption{(a) Temporal evolution of the magnetic, electric and particle kinetic energies integrated over the simulation domain. The left vertical axis refers to the variation of the magnetic ($\Delta \epsilon_B$) and particle kinetic ($\Delta \epsilon_K$) energies. The right vertical axis corresponds to the energy in the axial ($\epsilon_{Ez}$) and transverse ($\epsilon_{Ex} + \epsilon_{Ey}$) components  of the electric field. (b) Temporal evolution of the particle energy spectrum. The power-law tail extends beyond $\varepsilon_\mathrm{conf} = 125~m_ec^2$.
}
\end{center}
\end{figure}

The combination of electric and magnetic fields produced during the nonlinear stage of the KI is a potent accelerator of nonthermal particles. This is demonstrated by the formation of a nonthermal power-law tail in the particle energy spectrum, $dN / d\varepsilon \propto \varepsilon^{-p}$ with index $p = 2.75$ [Fig. 2(b)], that contains $\sim50\%$ of the initial magnetic energy. Moreover, the nonthermal component of the spectrum extends to the confinement energy $\varepsilon_{\rm conf} \equiv e B_0 R_c$, above which it rolls over as particles escape the system. The exceptional particles that attain the maximum energy do so by traveling along the jet axis $\hat z$ near the speed of light throughout the dynamical time of the KI; $\Delta \varepsilon_{\rm max} \simeq e \langle E_z \rangle c\tau_{\rm KI} \simeq 2 e B_0 R_c = 2 \varepsilon_{\rm conf}$. 
We have found that the power-law index decreases for weaker magnetizations, becoming $p \simeq 2$ for $\sigma \simeq 1$ (Supplemental Fig. S2).
This indicates that for $\sigma \simeq 1$, the highest energy particles can acquire a significant fraction of the jet's internal magnetic energy.

\begin{figure}[t!]
\begin{center}
\includegraphics[width=\linewidth]{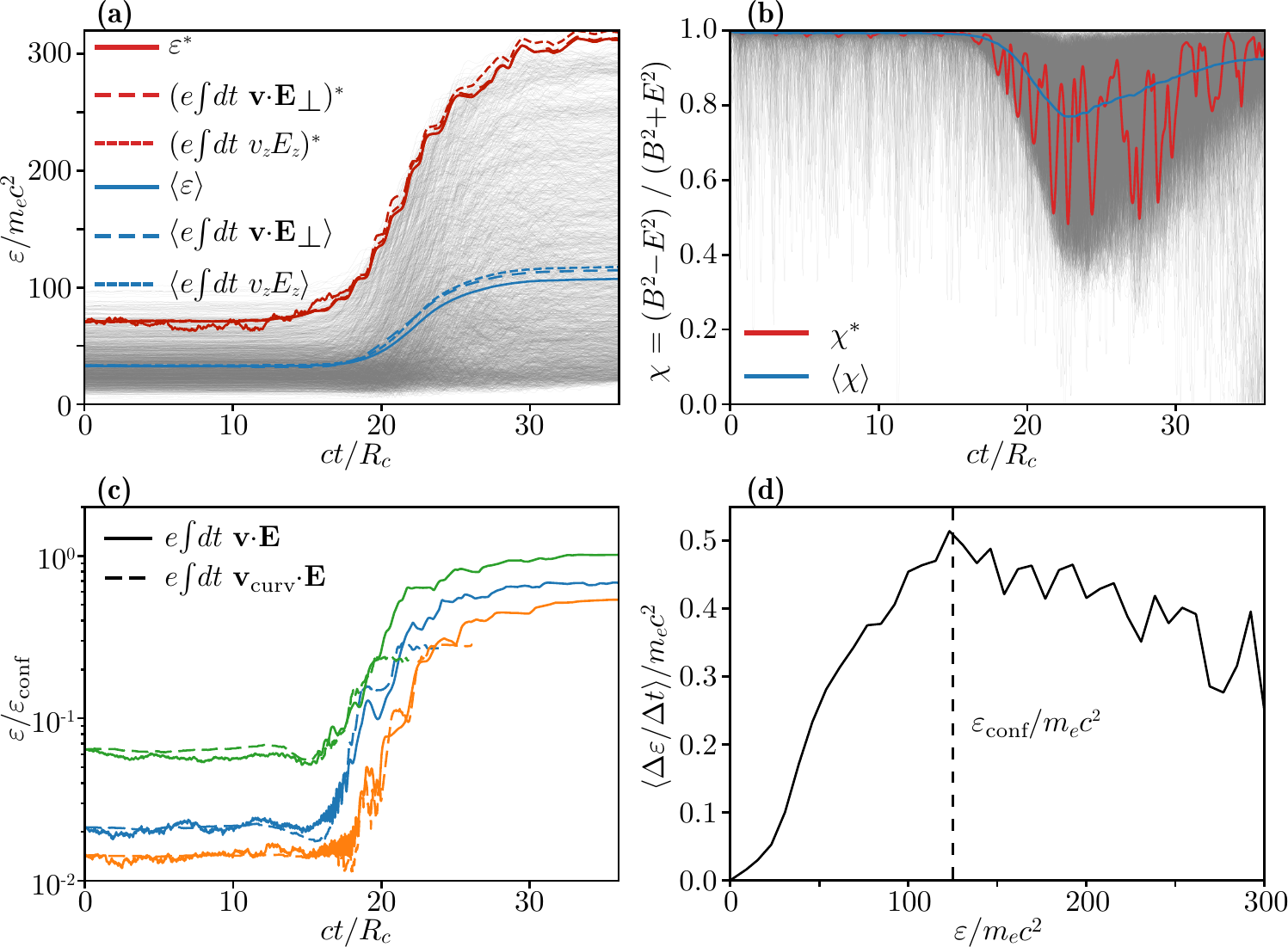}
\caption{(a) Evolution of particle energy $\varepsilon$ and (b) relative magnitude of $E$ and $B$ experienced by a representative sample of $2000$ nonthermal particles (grey curves). The red (blue) solid lines correspond to the highest (average) energy gain of the sample. The dashed and dotted curves indicate the integrated work done by $E_\perp$ and $E_z$, respectively, revealing that $E\simeq E_\perp \simeq E_z$ accounts for almost the entirety of the energy gain. (c) Evolution of the energy of 3 representative particles that start highly magnetized and reach $\varepsilon \sim \varepsilon_{\rm conf}$ (solid blue, orange, and green curves). Their acceleration is enabled by the curvature drift motion (dashed curves; these curves are interrupted when the guiding center description breaks down). (d) Mean energization rate, $\langle \Delta\varepsilon/\Delta t\rangle$, as a function of energy, $\varepsilon$, experienced by the $2000$ nonthermal particles during the period $18<ct/R_c<30$. The vertical dashed line corresponds $\varepsilon_{\rm{conf}}$.
}
\end{center}
\end{figure}

In order to uncover the mechanism responsible for the observed particle acceleration, we have performed a detailed analysis of the trajectories and the electromagnetic fields experienced by a representative sample of nonthermal particles \cite{suppl}.
Importantly, we find that these particles are primarily energized by the inductive electric field, $\mathbf{E} = -\mathbf{v} \times \mathbf{B} \simeq E_z \mathbf{\hat{z}}$, which implies $|\mathbf{E}| < |\mathbf{B}|$ [Fig. 3(a),(b)]. Nonideal, or parallel ($\mathbf E \parallel \mathbf B$) electric fields, commonly associated with the reconnection of magnetic field lines \cite{Sironi2014}, are thus not responsible for the observed particle acceleration.

Acceleration by an inductive electric field $\mathbf E \perp \mathbf B$ requires that particles cross magnetic field lines. This strongly suggests that magnetic field inhomogeneities must be crucial in the acceleration process. Indeed, our simulations reveal that the magnetic field becomes highly tangled, developing a spectrum of fluctuations over the complete range of scales, from $R_c$ down to the gyroradius scale $\langle \rho_g \rangle$ of thermal particles (Supplemental Fig. S3). We find that these fluctuations enable rapid displacement of particles across magnetic field lines via the guiding center curvature drift with velocity $\mathbf{v}_{\rm curv} = \gamma m v_\parallel^2 c \ \mathbf{B} \times \boldsymbol{\kappa} / eB^2$ \cite{Northrop1963} [Fig. (3c)]; $v_\parallel$ is the particle velocity parallel to the local magnetic field and $\boldsymbol{\kappa} \equiv \mathbf{B} \cdot \nabla \mathbf{B} / B^2$ is the magnetic field curvature vector. These drifts allow particles to gain energy from the inductive electric field, with the instantaneous rate of energy gain being connected to the spatial and temporal distribution of the magnetic field curvature. Fig. 3(c) illustrates the energy evolution of particles that start highly magnetized and reach the confinement energy. These particles experience fast energy gains by encountering regions where the field curvature radius $\kappa^{-1}$ is only a few times their gyroradius. Once they are accelerated to a large fraction ($0.2-0.3$) of the confinement energy, the guiding center approximation breaks down [Fig. (3c)], and particles become effectively unmagnetized, moving with $v_z \sim c$ along the jet axis. The acceleration stops when the particles either escape the jet spine in the transverse direction or when the electric field decays as the instability subsides.

The mean energization rate $\langle \Delta\varepsilon / \Delta t \rangle$ increases with $\varepsilon$ until $\varepsilon \lesssim \varepsilon_{\rm conf}$ [Fig. 3(d)], indicating a first-order Fermi process. This is consistent with acceleration via the curvature drift (since $v_{\rm curv} \propto \gamma$), the importance of which has also been identified in simulations of reconnecting current layers \cite{Dahlin2014,Guo2014}. We have confirmed that the same acceleration physics takes place in simulations with nonzero poloidal magnetic field, $B_z \lesssim 0.5 B_\phi$, yielding similar power-law particle energy spectra as the purely toroidal field case (Supplemental Fig. S4).

\begin{figure}
\begin{center}
\includegraphics[width=\linewidth]{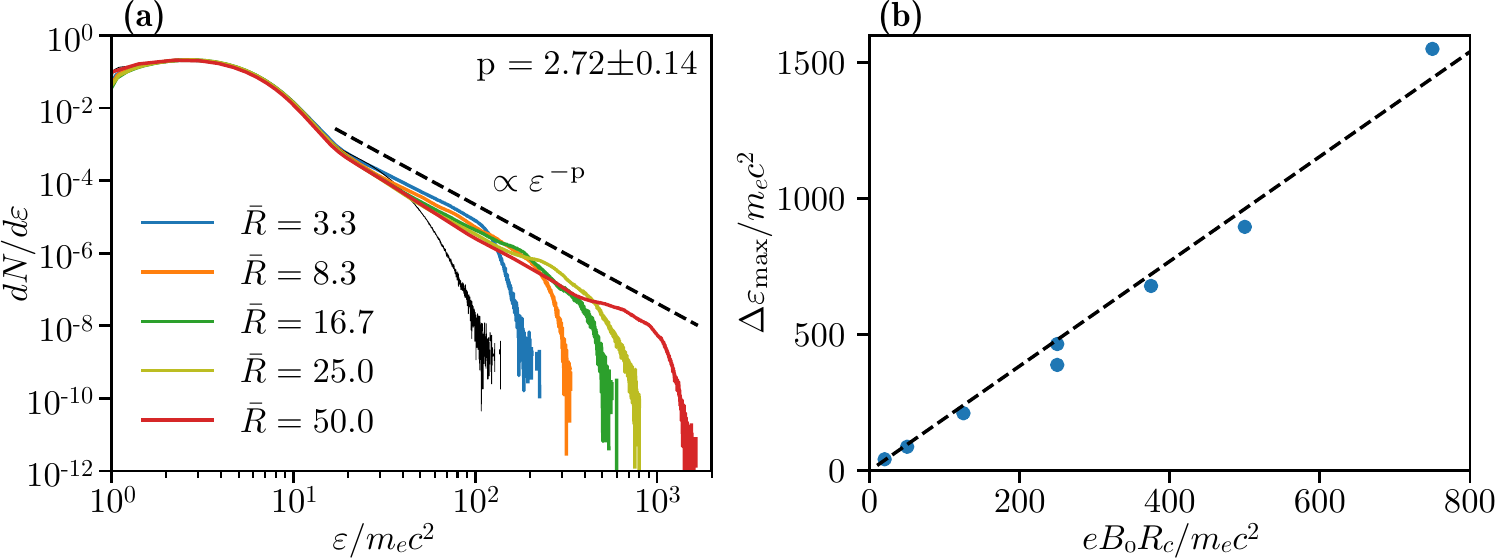}
\caption{(a) Final particle spectra for different system sizes ($\bar{R}$) and fixed magnetization $\sigma = 5$. The black curve corresponds to initial particle spectrum, which is the same for all system sizes.
(b) Scaling law of maximum particle energy gain with system size and magnetic field magnitude. The dots represent the results of 3D PIC simulations for different $\sigma$ and $\bar{R}$, and the dashed line represents the linear fit $\Delta \varepsilon_{\rm max} m_ec^2 \simeq 1.9~\varepsilon_{\rm conf} = 1.9~eB_0R_c$.
}
\end{center}
\end{figure}

The acceleration mechanism unveiled here is thus a consequence of a large-scale ($\sim R_c$) inductive electric field acting in concert with a magnetic field that is tangled over a range of scales that extends down to $\langle \rho_g \rangle$. These conditions are intrinsically 3D and arise self-consistently via the dynamic evolution of the KI \cite{suppl}. While the scales of $R_c$ and $\rho_g$ are vastly disparate in astrophysical jets, a small fraction of the initially thermal particles can always attain the confinement energy $\varepsilon_{\rm conf}$. We have confirmed this by systematically increasing the ratio $\bar{R}$ of the jet radius to the gyroradius of the thermal particles. The power-law spectral index is preserved, consistently extending from the thermal mean up to $\varepsilon_{\rm conf}$, and the maximum particle energy gain is always $\Delta \varepsilon_{\rm max} \simeq 2 e B_0 R_c$ (Fig. 4). This indicates that our results may be extrapolated to astrophysical systems, where the enormous scale separation implies huge energy gains.

Particle acceleration enabled by the KI can account for the synchrotron and inverse Compton (IC) radiating electrons in AGN jets. For example, the synchrotron energy spectrum, $F_\nu \propto \nu^{-\alpha}$, of knots such as HST-1 in M87 extends to hard X-rays, with spectral index $\alpha \simeq 1.0 - 1.5$ \cite{Harris2003, Harris2006}. The balance between synchrotron cooling and electron acceleration by the KI will steepen the electron spectrum, such that $\alpha$ is related to the spectral index $p$ of the injected particles by $\alpha = p / 2$ \cite{suppl}. This is in agreement with our results that show $p = 2 - 3$ for magnetization levels $\sigma = 1 - 10$ (Supplemental Fig. S2). The same population of relativistic electrons produces IC radiation at GeV - TeV photon energies, by up-scattering its own synchrotron (or ambient) photons \cite{PotterAndCotter2012}.

Protons and heavier ions are accelerated in the kink-unstable region of AGN jets in a similarly efficient way. This has been confirmed in simulations (Supplemental Fig. S5) with plasma having $1$ proton per $10$ electrons, the composition that is inferred by calorimetric modeling of many giant radio lobes \cite{Pjanka2016}. Unlike electrons, proton synchrotron losses are negligible up to and beyond the confinement energy. In HST-1, whose intrinsic scale is $R_c \sim \unit[]{pc}$ and magnetic field is $B_0 \sim 1-10$ mG, protons could attain energies $\unit[10^{18} - 10^{19}]{eV}$ by residing in the accelerating region throughout the $\tau_{KI} \sim \unit[30]{yr}$ (co-moving) dynamical time of KI. The brightest feature of M87, Knot A \cite{Stawarz2005}, has a significantly larger intrinsic scale $R_c \sim \unit[100]{pc}$ and may also possess mG-level magnetic fields. Under such conditions, the mechanism uncovered in this work would accelerate protons and iron nuclei to energies $\sim \unit[10^{20}]{eV}$ and $\sim \unit[10^{21}]{eV}$ respectively, with these most energetic particles acquiring a significant fraction of the jet's internal energy.
 
In summary, we have uncovered an efficient and robust particle acceleration mechanism that operates on the helical magnetic field structure of relativistic jets.  It has long been appreciated \cite{Hillas1984} that in AGN jets, $\varepsilon_{\rm conf}$ may exceed $10^{20}$ eV, which has made them prime UHECR source candidates. We have shown that charged particles can be efficiently accelerated to this limiting energy as a consequence of the internal KI, thus revealing a specific mechanism by which AGN jets could accelerate UHECRs. Our work also sheds new light on the generation of high-energy radiating particles in astrophysical jets and on the dependence of the radiation spectral index on the jet magnetization. We note that this mechanism can operate similarly in other astrophysical environments. Important examples are the Crab and other pulsar wind nebulae, for which the formation of a kink-unstable plasma column at high latitudes is observed and is expected to be a primary site where the pulsar's magnetic energy is dissipated \cite{Begelman1994,Lyubarsky2012,Porth2013,Zrake2017}.
 
The authors thank G. Madejski and R. Blandford for helpful discussions. This work was supported by the U.S. Department of Energy SLAC Contract No. DE-AC02-76SF00515, by the U.S. DOE Office of Science, Fusion Energy Sciences under FWP 100237, and by the U.S. DOE Early Career Research Program under FWP 100331. The authors acknowledge the OSIRIS Consortium, consisting of UCLA and IST (Portugal) for the use of the OSIRIS 3.0 framework and the visXD framework. Simulations were run on Mira (ALCF) through an ALCC award.
 
\bibliography{scibib.bib}
\bibliographystyle{apsrev4-1.bst}

\end{document}


\title{
Efficient Nonthermal Particle Acceleration by the Kink Instability in Relativistic Jets \\
Supplemental Material
} 


\author{E. P. Alves}
\email{epalves@slac.stanford.edu}
\affiliation{
High Energy Density Science Division, SLAC National Accelerator Laboratory, Menlo Park, CA 94025, USA
}%
\author{J. Zrake}
\affiliation{
Columbia Astrophysics Laboratory, Columbia University, 116th St \& Broadway, New York, NY 10027, USA
}%

\author{F. Fiuza}
\email{fiuza@slac.stanford.edu}
\affiliation{
High Energy Density Science Division, SLAC National Accelerator Laboratory, Menlo Park, CA 94025, USA
}%


\date{}





\maketitle

\section*{Simulation setup}
We model the self-consistent development of the KI in conditions relevant to relativistic astrophysical jets. In particular, our simulations capture i) 3D helical magnetic field geometries, and ii) relativistic magnetization regimes, \emph{i.e.} where the ratio $\sigma$ of magnetic to plasma rest-mass energy densities exceeds unity. Both of these conditions are central aspects of the magnetic field structure of astrophysical jets at $\sim \unit[100]{pc}$ propagation distances from the source. 
We therefore consider helical magnetic field configurations of the form $\mathbf{B}(r) = B_0\frac{r}{R_c}e^{1-r/R_c}\mathbf{e}_\phi + B_z \mathbf{e}_z$, where $R_c$ is the cross-sectional radius of the jet's kink-unstable inner core, and $B_0$ and $B_z$ are the amplitudes of the toroidal and poloidal components of the magnetic field component. The plasma current density $\mathbf{J} = \frac{c}{4\pi} \nabla \times \mathbf{B}$ is supported by symmetrically streaming electrons and positrons, corresponding to zero bulk flow of the pair plasma. Our simulations follow the evolution of the jet's kink unstable region in its co-moving frame. We consider a magnetostatic equilibrium configuration where the plasma thermal pressure ($P$) balances the hoop stress of the magnetic field; $\mathbf{J} \times \mathbf{B} - \nabla P = 0$. This configuration approximates a section of the jet plasma that was focused by recollimation and has stagnated due to its own thermal pressure, becoming highly vulnerable to the KI. We choose the plasma number density profile to be of the form $n(r) = n_0 + (n_c-n_0)/\mathrm{cosh}^2(2r/R_c)$, where $n_0$ and $n_c$ are the background and jet core densities, respectively, and the particle distributions are initialized according to drifting Maxwell-J\"uettner distributions.

Our choice to model the KI starting from a hydromagnetic equilibrium condition means that we do not capture certain features of the jet flow, such as velocity shear $v_z(r)$, expansion $v_r(r)$, or rotation $v_\phi(r)$. These features could in principle affect the development of the kink instability. For example, velocity shear may lead to Kelvin-Helmholtz type instabilities, which can compete with the KI. However, in the strongly magnetized regions considered in our work, the characteristic shearing velocity is sub-Alfv\'enic, and current-driven instabilities such as KI are expected to dominate. Radial expansion of the jet could have a stabilizing effect on the KI. However, one can show that the spine of magnetized jets remains highly collimated \cite{Lyubarsky2009}, and global MHD simulations \cite{Bromberg2016} confirm that the KI indeed develops at recollimation sites. This justifies the use of a cylindrical equilibrium configuration.

\sloppy Our initial unstable equilibrium is characterized by the dimensionless parameters $\sigma \equiv B_0^2/4\pi n_c m_e c^2$, $B_z/B_\phi \equiv B_z/B_0$, and $\bar{R} \equiv R_c/\langle \rho_g \rangle$, where $\langle \rho_g \rangle$ is the mean particle gyroradius in the core of the jet ($r\leq R_c$); $e$ and $m_e$ are the elementary charge and electron mass, respectively.
The dimensions of the simulated domains are $20\times20\times(10-20)~R_c^3$, with the jet placed at the center of the box and oriented along $\hat{\mathbf{z}}$. The grid resolution is set to $4-12$ points per $\langle \rho_g \rangle$ (corresponding to $3-8$ points per relativistic plasma skin depth at the core of the jet). Different resolutions were tested to ensure numerical convergence. We use $8-16$ particles per cell per species with quadratic particle shapes for improved numerical accuracy. No external perturbation is induced; the instability grows from thermal fluctuations. Periodic boundary conditions are imposed in all directions, and the simulation terminates before the accelerated particles cross the transverse boundaries. By the end of the simulations, the nonthermal component of the particle spectrum is no longer evolving.

In order to verify that our results do not depend sensitively on our choice of radial magnetic field profile $B_\phi (r)$, we have performed simulations with $B_\phi(r) \propto 1/r$ and $1/r^2$ and have verified that the overall dynamics of the KI and resulting non-thermal particle acceleration proceeds similarly to the case of the exponentially decaying profile. Our primary findings are thus not sensitive to the particular choice of magnetic field profile $B_\phi(r)$. We choose the exponentially decaying profile because it is most computationally economical.


\begin{figure}[t!]
\begin{center}
\includegraphics[width=\linewidth]{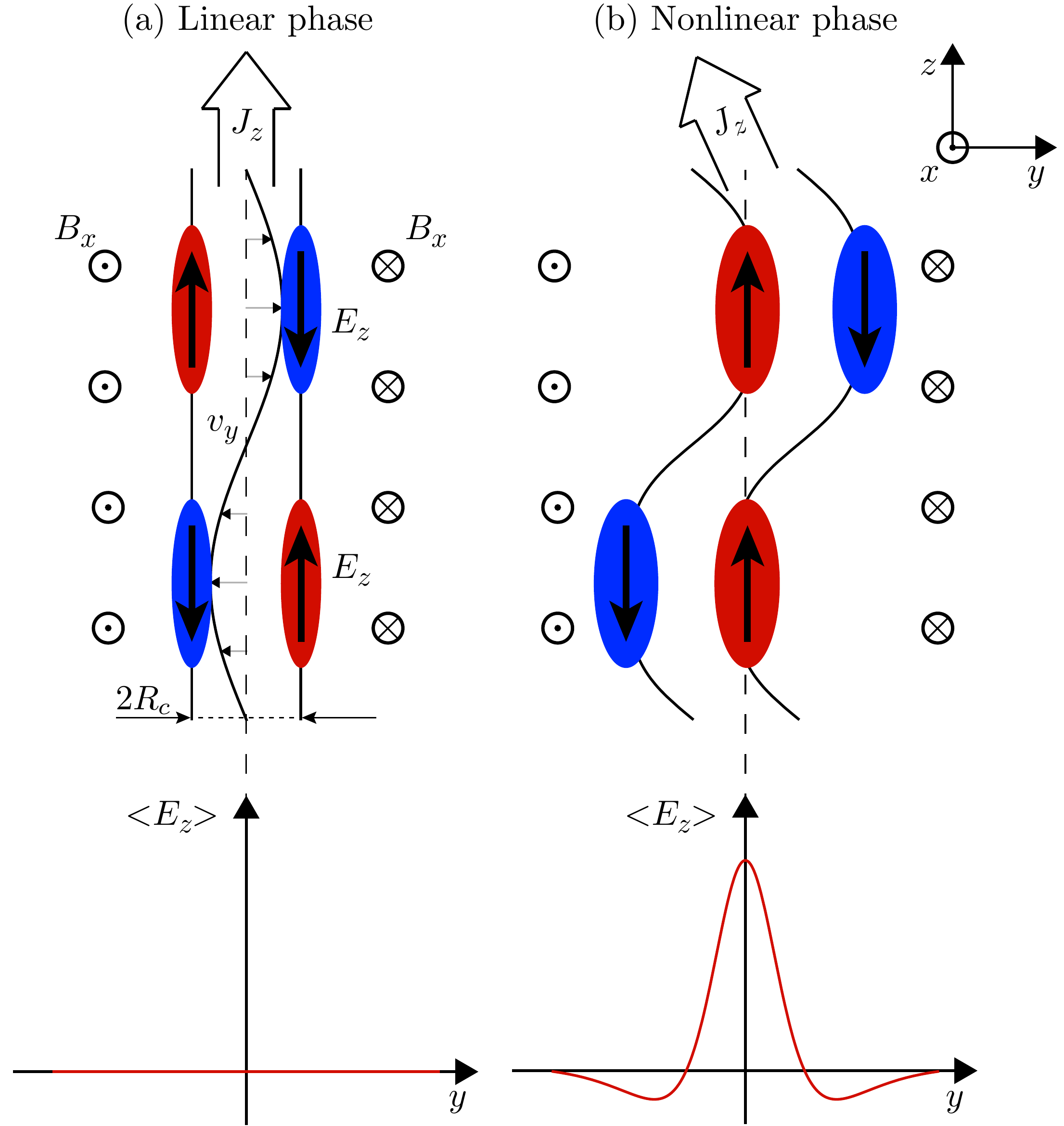}
\caption{Schematic illustration of the formation of a coherent inductive electric field along jet axis.
A cross section of the cylindrical jet current and the associated field structure is illustrated in the linear (a) and nonlinear (b) phases of the KI. The transverse distribution of the net axial electric field is presented in the lower plots.
}
\end{center}
\end{figure}

\section*{Formation of a coherent inductive electric field along jet axis}
We have shown in Fig. 1(c2) of the main text that a net axial electric field $\langle E_z\rangle$ emerges as a consequence of the nonlinear development of the KI. This component of the electric field (in concert with the highly tangled magnetic field) was shown to play a dominant role in the acceleration of nonthermal particles [Fig. 3(a) of the main text]. An illustrative schematic of how this electric field structure emerges from the development of the KI is presented in Fig. S1. For simplicity, we illustrate this effect in the 2D $yz$ plane passing at the center of the jet (the coordinate system is presented in the top right of Fig. S1). Here, the toroidal magnetic field is given by $B_x$, which points in/out of the $yz$ plane. The onset of the KI induces radial motions of the jet that will ultimately result in the characteristic kink distortions. The fluid velocity associated with these radial motions is illustrated in Fig. S1(a), projected onto the $yz$ plane, $v_y$; the velocity $v_y$ is illustrated as a harmonic perturbation along the axis of the jet, corresponding to an unstable eigenmode of the KI. This motion induces the electric field $\mathbf{E} = -\mathbf{v}\times\mathbf{B} \Rightarrow E_z = v_y B_x$ , which is represented by the colored ellipses at the edges of the jet; red/blue colors represent positive/negative regions of $E_z$. At early times (in the linear phase of the KI), the kink distortions of the jet are negligible, and thus the axially averaged electric field $\langle E_z\rangle$ is essentially zero, i.e. there is no net axial electric field. However, as the kink distortions grow and the instability enters the nonlinear regime, the distortions bring into alignment regions of like-oriented $E_z$ (the red ellipsoidal regions are brought into alignment at the center of the jet). This gives rise to a net axial electric field as shown in Fig. S1(b). This electric field persists during the transit time of the kink perturbations across the diameter of the jet, which corresponds to $\sim 10R_c/c$.

\section*{Dependence of nonthermal power-law index $p$ on magnetization $\sigma$}
We observe that the index $p$ of the nonthermal power-law tail varies with magnetization $\sigma$ of the jet. This is illustrated in Fig. S2, which shows the measured $p$ indices from the set of simulations performed in this work for different magnetizations. The error bars indicate the standard deviation of the index measurements from simulations of jets with fixed $\sigma$ and varying scale size $\bar{R}$.
We find that the index $p$ increases from $2-3$ with increasing $\sigma$ in the range $1-10$. Thus, for magnetizations $\sigma \sim 1$, a large fraction of the dissipated magnetic energy is transfered to the highest energy particles.

\begin{figure}[b!]
\begin{center}
\includegraphics[width=0.925\linewidth]{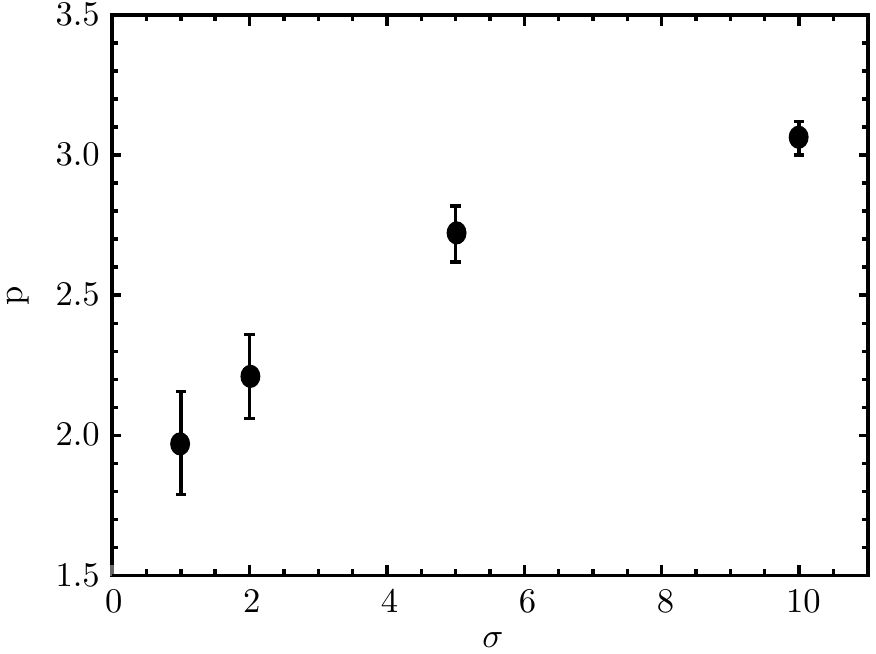}
\caption{Dependence of nonthermal power-law index $p$ on magnetization $\sigma$. Measured nonthermal indexes $p$ from the set of simulations performed in this work for different $\sigma$ and $\bar{R}$.
}
\end{center}
\end{figure}

\section*{Spectrum of magnetic fluctuations}
As discussed in the main text, the accelerating electric field is inductive, \emph{i.e.} $\mathbf{E}\perp\mathbf{B}$. This means that magnetic field inhomogeneities must be present to allow particles to displace across magnetic field lines so that they can be accelerated by the inductive electric field. Indeed we find that the jet's magnetic field structure develops a spectrum of spatial fluctuations as a consequence of the KI (Fig. S3). The spectra in Fig. S3 are obtained by taking the Fourier transform of $B_x$ and $B_y$, squaring the amplitudes, and averaging over $k_x$ and $k_y$ (the transverse wave numbers). This figure shows that while there is a well-defined wavenumber in the linear phase of the instability (associated with the fastest growing helical kink mode, $k_z R_c \simeq 2.5$), the nonlinear phase of the instability is characterized by a spectrum of fluctuations that extends down to the gyroradius of thermal particles $\langle\rho_g\rangle$. These spectral features of the magnetic fluctuations were consistently found in all simulations performed in this work.

\begin{figure}[t!]
\begin{center}
\includegraphics[width=\linewidth]{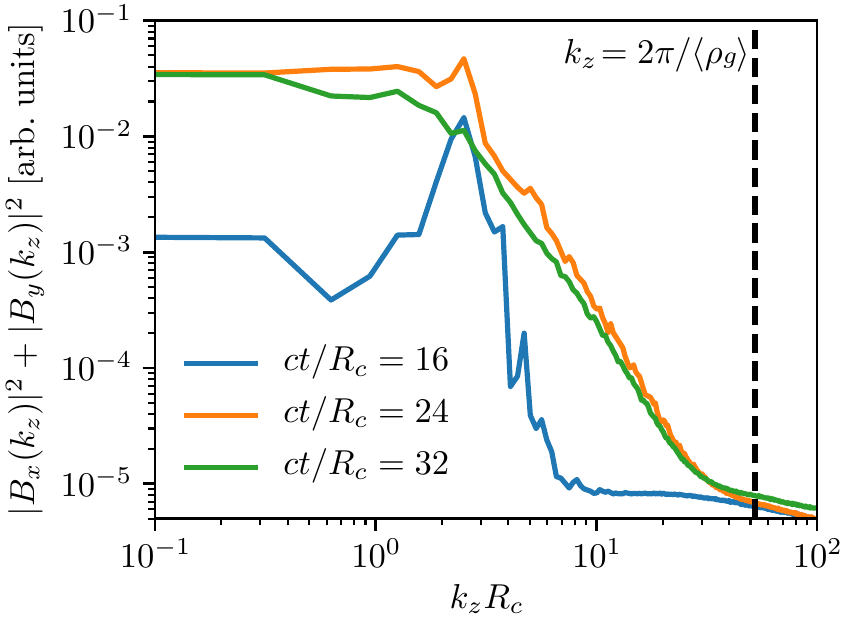}
\caption{Evolution of the power spectrum of magnetic field fluctuations for the case $\bar{R} = 8.3$ and $\sigma = 5$.
The spectrum of spatial fluctuations of the magnetic field along the jet axis, $k_z$, is shown at times $ct/R_c = 16$, $24$, and $32$. The vertical dashed line marks the wavenumber corresponding to the $2\pi/\langle\rho_g\rangle$, \emph{i.e.}, spatial modulations at the scale of the gyroradius of thermal particles in the spine of the jet.}
\end{center}
\end{figure}

\section*{Particle selection and acceleration analysis}
The particle trajectories presented in Fig. 3 of the main text are representative of the acceleration dynamics of nonthermal particles produced by the development of the KI in our simulations. This representative particle sample was obtained by randomly selecting $100$ particles from $20$ logarithmically-spaced energy bins from the nonthermal component of the final particle energy spectrum.

The plots presented in Fig. 3(d) were obtained from the particle trajectories during the acceleration phase $18<ct/R_c<30$. During this period, the time-history of each particle trajectory was divided into $10$ segments of duration $T=1.2~ct/R_c$, that represent isolated acceleration events. Note that we have verified that our results are insensitive to duration $T$ of the trajectory segments. The quantities $\Delta\varepsilon/\Delta t \equiv (\varepsilon_f-\varepsilon_0)/T$ and $\langle\varepsilon \rangle$ were computed for each segment; $\langle Q \rangle = (1/T) \int_0^T Q(t)\mathrm{d}t$ denote time averages of quantity $Q$ for each segment. We then computed the probability density function $P(\Delta\varepsilon/\Delta t | \langle \varepsilon \rangle)$ (the probability of an acceleration event yielding $\Delta\varepsilon/\Delta t$ given that the particle had kinetic energy $\langle \varepsilon \rangle$). The curve in 3(d) was obtained by taking the first moment in $\langle\Delta\varepsilon/\Delta t \rangle$ of each of this probability distribution function:

\begin{equation}
\begin{split}
\langle\Delta & \varepsilon/\Delta t \rangle(\langle \varepsilon \rangle) = 
\\
&\int {\Delta\varepsilon/\Delta t ~ P(\Delta\varepsilon/\Delta t | \langle \varepsilon \rangle)~\mathrm{d}(\Delta\varepsilon/\Delta t)}.
\end{split}
\end{equation}

\section*{Role of a finite poloidal magnetic field component on particle acceleration dynamics}
We have explored the role of a finite poloidal magnetic field component ($B_z$) on the particle acceleration dynamics. It is well known that the tension of a poloidal magnetic field component has a stabilizing effect on the KI \cite{Bateman1978}. For sufficiently strong $B_z$, the KI can even be fully suppressed. If the poloidal component is not too large as to fully suppress the KI ($B_z < B_\phi$), the growth rate of the instability is lowered and the fastest growing mode is modified. Interestingly, however, we continue to observe efficient particle acceleration with similar characteristics as the purely toroidal case (Fig. S4).

We have simulated the jet with $\bar{R} = 8.3$, $\sigma = 5$, and a spatially uniform poloidal magnetic field for three different strengths, $B_z/B_0 = 0.0$, $0.1$ and $0.5$. (Note that the toroidal component of the magnetic field is kept fixed for this set of simulations.) Figure S4 shows the spectrum of accelerated particles for the three cases. For a weak poloidal component ($B_z/B_\phi = 0.1$), as expected in the recollimation regions of AGN jets at $\sim \unit[100]{pc}$ distances from the central engine, we observe efficient acceleration of nonthermal particles with identical properties as the purely toroidal case. When the poloidal and toroidal components are comparable ($B_z/B_\phi = 0.5$), the growth rate of the instability and the amplitude of the inductive electric field start to become smaller. However, even in this case, a similar nonthermal particle spectrum is developed, with the maximum energy reaching the confinement energy of the jet. Note that the uniform poloidal magnetic field component is not dissipated by the development of the KI due to flux conservation, and hence its energy is not converted into particle energy.

When the magnetic field has a poloidal component, the electric field induced by the KI develops also off-axis components. In this case, the electric field has a helical structure. We have verified that this does not change the underlying particle acceleration mechanism: i) the acceleration is still dictated by the inductive electric field ($\mathbf{E} = -\mathbf{v} \times \mathbf{B}$), and ii) is facilitated by highly tangled magnetic fields that enable fast curvature drifts across field lines, allowing particles to efficiently displace along the inductive electric field and gain energy. We have thus confirmed the main features of the acceleration mechanism described in our work continue to operate for $B_z/B_\phi \leq 0.5$.

\begin{figure}[t!]
\begin{center}
\includegraphics[width=\linewidth]{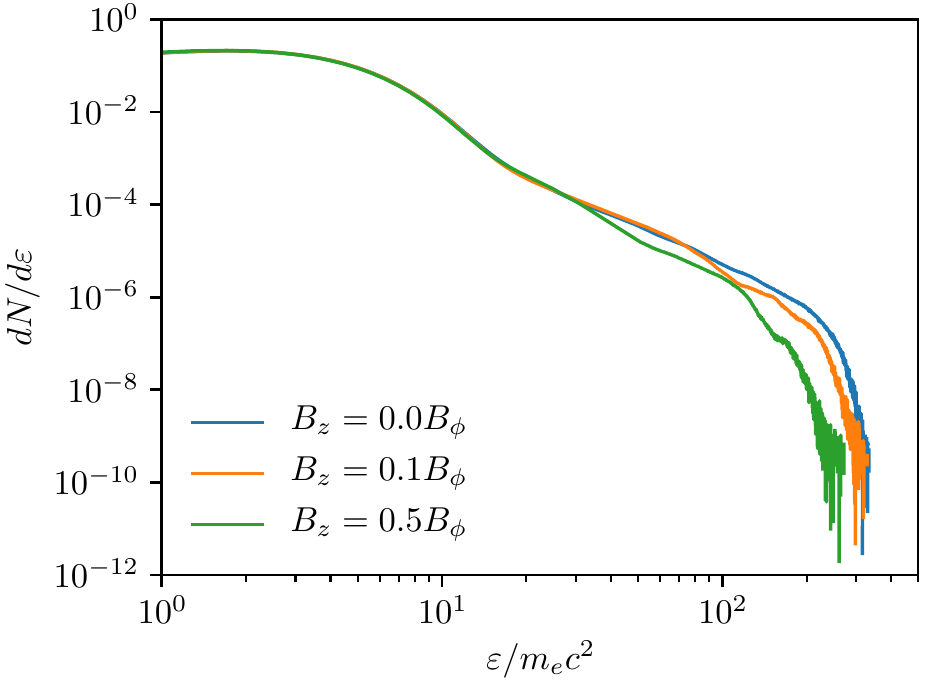}
\caption{Robustness of particle acceleration mechanism to the presence of a finite poloidal magnetic field component ($\mathbf{B_z}$).
The magnetic field structure has the same toroidal component ($B_\phi$) in all simulations, but a uniform poloidal component with different amplitudes $B_z = 0.0 B_\phi$ (blue curve), $0.1 B_\phi$ (orange curve) and $0.5 B_\phi$ (green curve). In all cases, nonthermal particles attain the confinement energy $\varepsilon_\mathrm{conf} = 125~m_ec^2$.
}
\end{center}
\end{figure}

\section*{Importance of 3D simulations}
The particle acceleration mechanism we have observed in our simulations depends critically on the 3D dynamics and magnetic field geometry of the KI. We have performed 2D simulations (not presented here) of the corresponding 2D KI setup. We have observed no significant particle acceleration, consistent with previous studies of planar current layers \cite{Zenitani2005}. This inneficiency can be understood by the absence of in-plane curvature drifts in this 2D geometry. In addition, canonical momentum conservation out of the simulation plane limits particle transport across field lines \cite{Jokipii1993}.

\section*{Synchrotron radiation from nonthermal leptons}
In realistic astrophysical jets, where particles can accelerate to enormous energies, synchrotron cooling effects will strongly impact the spectrum of nonthermal electrons and positrons accelerated by the KI. The change in the nonthermal spectrum is expected to occur when the synchrotron cooling time becomes comparable to the dynamical time of the KI, i.e. 
\begin{equation}
\nonumber
\frac{\varepsilon}{d\varepsilon/dt\vert_\mathrm{cool}} = \tau_\mathrm{KI},
\end{equation}
where $d\varepsilon/dt\vert_\mathrm{cool} = 4/9~e^4/m_e^2c^3~B_0^2\gamma^2 \beta^2$ is the synchrotron cooling rate, $\gamma = \varepsilon / m_e c^2$ is the relativistic Lorentz factor of the particles, and $\beta = \sqrt{1 - 1 / \gamma^2}$ is the three velocity in units of $c$. The electron/positron energy at which this occurs is given by $\varepsilon_{\rm cool} \simeq 0.4 \mathrm{TeV}/B_\mathrm{mG}^2R_\mathrm{pc}$, where $B_\mathrm{mG}$ and $R_\mathrm{pc}$ are the nominal magnetic field amplitude in mG and the jet spine radius in parsec, respectively. Above this critical energy, synchrotron cooling effects become important and steepen the nonthermal power-law spectrum to $p+1$. This broken power-law distribution of the electron/positron energies is reflected in the resulting spectrum of  synchrotron emission. The synchrotron energy spectrum is given by $F_\nu \propto \nu^{-\alpha}$, with $\alpha = (p-1)/2$ below and $\alpha = p/2$ above the spectral break at $\nu_{\rm break} \simeq 10\mathrm{eV}/B_\mathrm{mG}^3R_\mathrm{pc}^2$.

For the HST-1 knot in M87, which is inferred to possess a characteristic scale $R_c\sim1$pc and a nominal magnetic field strength $B_0 \sim 1$mG, the synchrotron spectral break is expected to occur at $\nu_\mathrm{break} = 10\mathrm{eV}$, which lies in the UV range. For the relativistic magnetizations considered in our work $\sigma=1-10$, which produce nonthermal particles with index $p=2-3$ (Fig. S2), the synchrotron spectral indexes are expected to be $\alpha_\mathrm{RO} = 0.5-1$ and $\alpha_\mathrm{X} = 1-1.5$ for the radio-optical and X-ray spectral ranges, respectively. These are consistent with \emph{Hubble} and \emph{Chandra} observations of M87 \cite{Harris2003, Harris2006}.

Synchrotron losses experienced by electrons and positrons place a hard upper limit on their energy of $\unit[0.6]{PeV} B_\mathrm{mG}^{-1/2}$. This limiting energy is attained when the maximum acceleration rate $\Delta \varepsilon_{\rm max} / \tau_{\rm KI}$ balances the synchrotron loss rate $d\epsilon/dt\vert_\mathrm{cool}$. Electrons and positrons reach this limiting energy if they experience the maximum acceleration rate continuously over a period (in the jet's co-moving frame) of $\sim 4 B_{\rm \rm mG}^{-3/2}\unit[]{days}$. At this limiting energy, electrons radiate $\sim \unit[20]{MeV}$ $\gamma$-rays.

\begin{figure}[t!]
\begin{center}
\includegraphics[width=\linewidth]{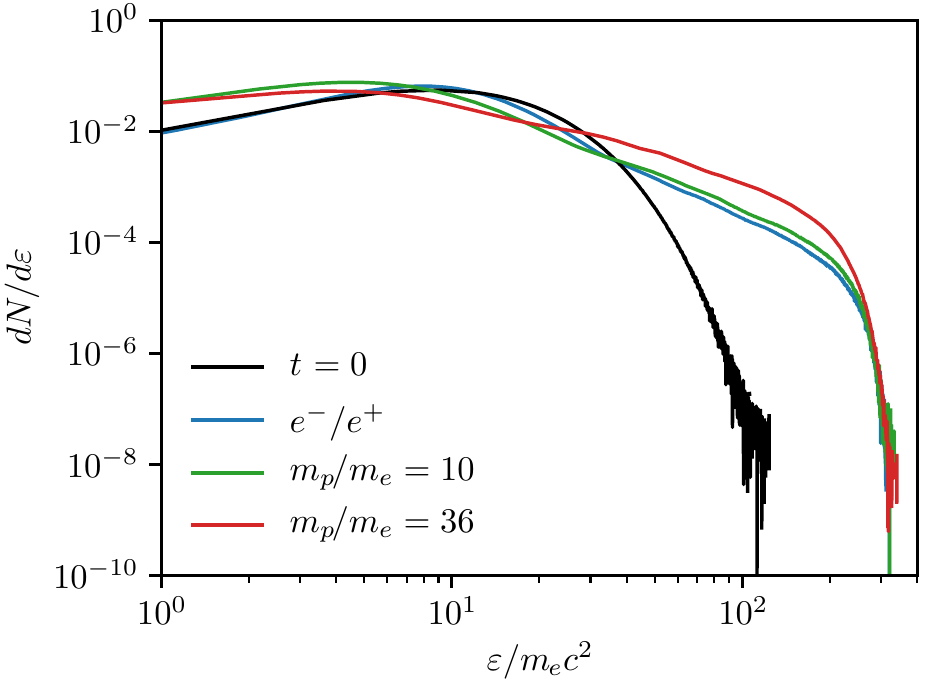}
\caption{Evidence of nonthermal acceleration of protons. Spectrum of accelerated particles that were initialized in the spine of the jet ($r\leq R_c$), for $\bar{R} = 8.3$, $\sigma=5$, and a jet composition with a $10\%$ ion fraction. The spectra for ions with $m_i/m_e = 10$ (green curve) and $36$ (red curve) are compared with the leptonic spectrum (blue curve), confirming that heavier particles also undergo similar nonthermal acceleration and that they attain the confinement energy $\varepsilon_\mathrm{conf} = 125~m_ec^2$.}
\end{center}
\end{figure}

\section*{Efficient nonthermal acceleration of protons}
In order to verify that protons can be accelerated similarly to positrons in the KI-induced electric and magnetic fields, we have performed simulations with mixed electron-positron-proton jet compositions. We have simulated a jet with $\bar{R} = 8.3$, $\sigma = 5$ and $B_z/B_\phi = 0$, and composition of $90\%$ positrons and $10\%$ protons. We have used reduced proton mass ratios of $m_p/m_e = 10$ and $36$, which allow for scale separation between protons and positrons, within a reasonable increase in computational size. The protons were initialized in thermodynamic equilibrium with the electrons and positrons.

The accelerated proton spectrum in Fig. S5 confirms that protons undergo similar nonthermal acceleration dynamics, exhibiting a power law tail that extends to the confinement energy $\varepsilon_{\rm conf}$. This can be understood on the basis that positrons and protons of similar energy experience similar curvature drift motions and thus are similarly accelerated by the inductive electric field. Note that the slight differences between the spectral curves at energies below $m_p c^2$ are associated with the transition from nonrelativistic to relativistic particle energies. At relativistic energies, the energy spectra of all plasma species are mutually alike. These results thus indicate that electrons, positrons and protons are similarly accelerated via the development of the KI. Synchrotron losses do not limit the energization of protons and heavier ions as they do for leptons.

\bibliography{scibib}
\bibliographystyle{apsrev4-1.bst}